# Developing an Analytical Model of Frequency and Voltage Variations for Dynamic Reconfiguration

Jae-Young Park, *Student Member, IEEE* and Young-Jin Kim, *Member, IEEE*

*Abstract*—This paper develops a new analytical model to estimate real-time variations in grid frequency and voltages resulting from dynamic network reconfiguration (DNR). In the proposed model, switching operations are considered as discrete variations in an admittance matrix, leading to step variations in node injection currents. The network model with discrete admittance variations is then integrated with dynamic models of synchronous generators and voltage-dependent loads, enabling analysis of the dynamic grid operations initiated by the DNR. Case studies are performed to validate the proposed model via comparison with a conventional model and a comprehensive MATLAB/SIMULINK model.

## NOMENCLATURE

| | |
|---|---|
| $i, j, k$ | indices for nodes |
| $g$ | index for generators |
| $B, A$ | subscripts for network condition before and after DNR |
| $N, N_G$ | numbers of nodes and generators |
| $H_g$ | moment of inertia of generator $g$ |
| $f, \mathbf{F_{SG}}$ | grid frequency and rotational frequencies of generators |
| $G_{ij}, B_{ij}$ | real and imaginary parts of line admittance between nodes |
| $\mathbf{Y}, \mathbf{y}_{ij}$ | admittance matrices for network and line between nodes $i$ and $j$ |
| $\mathbf{Z}$ | impedance matrix with shunt components of distribution lines, generators, and loads |
| $\mathbf{I}, \mathbf{V}$ | $dq$-axis node injection currents and voltages |
| $|\mathbf{V}|$ | node voltage magnitude |
| $\mathbf{Y_L}$ | coefficient matrix for aggregated load model |
| $\mathbf{A_{SG}}, \mathbf{B_{SG}}, \mathbf{C_{SG}}, \mathbf{Y_{SG}}$ | coefficient matrices for aggregated generator model |
| $\mathbf{C_F}, \mathbf{C_V}, \mathbf{D_V}$ | coefficient matrices for the model outputs $\Delta f$ and $\Delta|\mathbf{V}|$ of the power grid including synchronous generators and loads |
| $\mathbf{X_{SG}}, \mathbf{X}$ | state variables for generators and power grid |
| $\mathbf{M}$ | matrix to convert $\Delta \mathbf{V}$ to $\Delta|\mathbf{V}|$ |
| $\mathbf{S}$ | matrix to extract $\Delta \mathbf{F_{SG}}$ from $\Delta \mathbf{X}$ |

## I. INTRODUCTION

DYNAMIC network reconfiguration (DNR) is the process of altering the topological structure of a power grid by opening and closing the sectionalizing and tie switches [1]. DNR often aims to accommodate the increased renewable power generation and reduce network power losses and outage durations. In many papers, optimal scheduling of reconfigurable switches, generators, and loads was studied to achieve these objectives in a steady state [1], [2].

The effects of DNR on real-time frequency and voltages have recently been considered only in a few studies [2], [3]. In [2], small signal analysis was conducted, considering the transient-free topology variation achieved by minimizing the voltage difference between the two sides of the switches in advance. In the conventional analytical model of [3], the DNR was considered as the variation in total load demand according to the amount to be restored or shed. To our knowledge, there has been no analytical model that enables analyzing the effects of DNR on real-time variations in node voltages, line power losses, and grid frequency. Instead, the analysis has relied on simulation models of reconfigurable power grids that are implemented using software such as MATLAB/SIMULINK.

This paper develops a new analytical model to analyze the effects of DNR on real-time grid operation, as shown in Fig. 1. In the proposed model, the DNR (i.e., the switch operations) is reflected as discrete variations in line admittances, which characterize the relationship between $dq$-axis node currents and voltages. These discrete variations lead to step variations in the node currents that arise when the network topology changes. The network model based on the discrete admittance variations is then combined with dynamic models of synchronous generators (SGs) and voltage-dependent loads. This enables estimation of transient and steady-state variations in node voltages, line power losses, and grid frequency resulting from the DNR. Case studies are performed to validate the proposed analytical model via comparison with a conventional model and a comprehensive MATLAB/SIMULINK model of the IEEE 37-node test feeder. The proposed DNR model is simple, and thus can be readily integrated with small signal models of power equipment, such as those discussed in [2] and [3], to support the development of control schemes and analysis of grid stability.

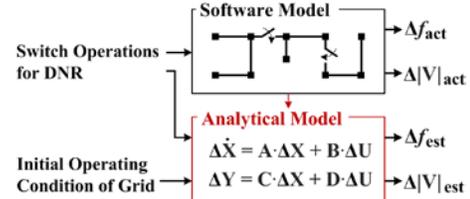

Fig. 1. Development of an analytical model for the grid response to DNR

## II. ANALYTICAL MODEL OF A RECONFIGURABLE GRID

### A. Modeling the Effects of DNR on Node Currents

For a steady-state operating condition, a bus injection model (BIM) of an electric network is represented as:

$$\mathbf{I_0} = \mathbf{Y_B} \cdot \mathbf{V_0}. \quad (1)$$

where $\mathbf{I_0}$ and $\mathbf{V_0}$ are the $dq$-axis currents injected to the nodes and the $dq$-axis voltages at the nodes, respectively. For example, $\mathbf{I_0}$ and $\mathbf{V_0}$ for node $i$ are defined as:

$$\mathbf{I_{0}}_i = [I_{di0}, I_{qi0}]^T \text{ and } \mathbf{V_{0}}_i = [V_{di0}, V_{qi0}]^T. \quad (2)$$

The admittance matrix $\mathbf{Y_B}$ in (1) consists of block matrices whose elements for the line between nodes $i$ and $j$ are shown as:

$$\mathbf{Y_{B}}_{ij} = \mathbf{y}_{ij} = \begin{bmatrix} G_{ij} & -B_{ij} \\ B_{ij} & G_{ij} \end{bmatrix}. \quad (3)$$

Note that for simplicity, a three-phase (3-*ph*) balanced network is considered for (2) and (3), where the 0-axis components are

J. Park and Y. Kim are with the Department of Electrical Engineering, Pohang University of Science and Technology, Pohang, Gyungbuk 790-784, Korea (e-mail: powersys@postech.ac.kr).

zero; the proposed model can still be applied to an unbalanced network. After the DNR is initiated, the BIM changes to:

$$\mathbf{I_0} + \Delta \mathbf{I} = \mathbf{Y_A} \cdot (\mathbf{V_0} + \Delta \mathbf{V}), \quad (4)$$

where $\Delta \mathbf{I}$ and $\Delta \mathbf{V}$ are the perturbations of the node currents and voltages, respectively, and $\mathbf{Y_A}$ is the admittance matrix for the reconfigured network. From (1) and (4), $\Delta \mathbf{I}$ is represented as:

$$\begin{aligned}\Delta \mathbf{I} &= \mathbf{Y_A} \cdot \Delta \mathbf{V} + (\mathbf{Y_A} - \mathbf{Y_B}) \cdot \mathbf{V_0} = \mathbf{Y_A} \cdot \Delta \mathbf{V} + \Delta \mathbf{Y} \cdot \mathbf{V_0}\\ &= \mathbf{Y_A} \cdot \Delta \mathbf{V} + \Delta \mathbf{I_T},\end{aligned} \quad (5)$$

where the DNR is reflected as the discrete variations in the admittance matrix $\Delta \mathbf{Y}$, leading to the step variations $\Delta \mathbf{I_T}$ that arise right immediately the network topology changes (i.e., $\mathbf{Y_B} \neq \mathbf{Y_A}$). In (5), $\Delta \mathbf{I_T}$ does not decay with time, driving the power grid to another steady state. In other words, $\Delta \mathbf{I_T}$ affects the network power losses and the power inputs of voltage-dependent loads in both transient and steady states. This shows the necessity of reflecting $\Delta \mathbf{I_T}$ in small-signal models for estimating the grid frequency and voltages.

To better understand $\Delta \mathbf{I_T}$, Fig. 2 shows an example of part of an $N$-node network with two reconfigurable switches, each of which is located between two adjacent nodes. The DNR is achieved by simultaneously closing $SW_{ij}$ and opening $SW_{jk}$, resulting in $\Delta \mathbf{y}_{ij} = \mathbf{y}_{ij}$ and $\Delta \mathbf{y}_{jk} = -\mathbf{y}_{jk}$. As shown in (6), $\Delta \mathbf{I_T}$ has non-zero elements only for the nodes linked via the switches; the corresponding values are determined based on the node voltages before the DNR and the discrete admittance variations.

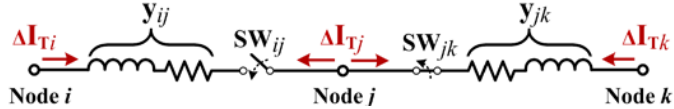

Fig. 2. An example of part of a reconfigurable network for $\Delta \mathbf{I_T}$ in (6).

B. *Dynamic Models of SGs and Voltage-dependent Loads*

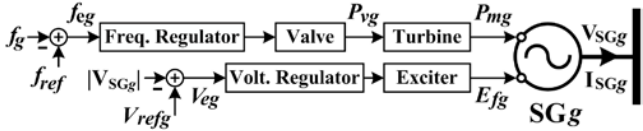

Fig. 3. Schematic diagram of a common dynamic model of an SG.

Commonly, SGs are used as grid-forming units that maintain the grid frequency and voltages in real time. Fig. 3 shows a schematic diagram of a dynamic model of an SG, where the frequency is regulated via droop and proportional-integral (PI) controllers and the terminal voltage is adjusted using a PI controller [4]. The dynamic model for all SG units $g = 1, \cdots, N_G$ can be represented in aggregated form as:

$$\Delta \dot{\mathbf{X}} = \mathbf{A_{SG}} \cdot \Delta \mathbf{X} + \mathbf{B_{SG}} \cdot \Delta \mathbf{V}, \quad (7)$$

$$\Delta \mathbf{I_{SG}} = \mathbf{C_{SG}} \cdot \Delta \mathbf{X} - \mathbf{Y_{SG}} \cdot \Delta \mathbf{V}, \quad (8)$$

$$\text{where} \quad \Delta \mathbf{X} = [\Delta \mathbf{X_{SG_1}}^T, \cdots, \Delta \mathbf{X_{SG_g}}^T, \cdots, \Delta \mathbf{X_{SG_{N_G}}}^T]^T. \quad (9)$$

In (7) and (8), $\mathbf{A_{SG}}$, $\mathbf{B_{SG}}$, $\mathbf{C_{SG}}$, and $\mathbf{Y_{SG}}$ are block diagonal matrices, each of which includes the linearized model parameters of an SG in a steady state. The element-wise expressions of the matrices are provided in [4]. In (9), $\Delta \mathbf{X_{SG_g}}$ is defined as a set of eight state variables, $[\Delta f_g, \Delta \delta_g, \Delta P_{mg}, \Delta P_{vg}, \int \Delta f_{eg}, \Delta E'_{qg}, \Delta E_{fg}, \int \Delta V_{eg}]^T$, for the third-order model of SG unit $g$ [5].

Furthermore, the ZIP load model [6] has been used widely to represent the static and dynamic responses of voltage-dependent loads. The aggregated model of all loads is:

$$\Delta \mathbf{I_L} = \mathbf{Y_L} \cdot \Delta \mathbf{V}, \quad (10)$$

where $\mathbf{Y_L}$ is a block diagonal matrix of the ZIP coefficients of the linearized load models. The element-wise expression of $\mathbf{Y_L}$ is provided in [4].

C. *Complete Model for Frequency and Voltage Estimation*

The proposed analytical model of a reconfigurable grid can be established by combining (5) with (7)–(10). Specifically, $\Delta \mathbf{V}$ is expressed with respect to $\Delta \mathbf{X}$ and $\Delta \mathbf{I_T}$ by substituting (8) and (10) into (5), given $\Delta \mathbf{I} = \Delta \mathbf{I_{SG}} + \Delta \mathbf{I_L}$, as:

$$\Delta \mathbf{V} = \mathbf{Z} \cdot (\mathbf{C_{SG}} \cdot \Delta \mathbf{X} + \Delta \mathbf{I_T}), \quad (11)$$

where $\mathbf{Z} = (\mathbf{Y_L} - \mathbf{Y_A} - \mathbf{Y_{SG}})^{-1}$. Using (7) and (11), the dynamic responses of the SGs and loads to the DNR are represented as:

$$\Delta \dot{\mathbf{X}} = \mathbf{A} \cdot \Delta \mathbf{X} + \mathbf{B} \cdot \Delta \mathbf{I_T}, \quad (12)$$

where $\mathbf{A} = \mathbf{A_{SG}} + \mathbf{B_{SG}} \cdot \mathbf{Z} \cdot \mathbf{C_{SG}}$ and $\mathbf{B} = \mathbf{B_{SG}} \cdot \mathbf{Z}$.

In addition, $\Delta f$ due to the DNR can be estimated using $\Delta \mathbf{X}$, based on the principle of the center of inertia [7], as:

$$\Delta f = \mathbf{H} \cdot \Delta \mathbf{F_{SG}} = \mathbf{H} \cdot \mathbf{S} \cdot \Delta \mathbf{X}, \quad (13)$$

$$\text{where} \quad \mathbf{H} = \left(\sum_{g=1}^{N_G} H_g\right)^{-1} \left[H_1, H_2, \cdots, H_{N_G}\right]. \quad (14)$$

In (13), $\mathbf{S}$ consists of $N_G$ row vectors, each of which is equal to $[1, \mathbf{O}_{7 \times 1}]$ for $\Delta \mathbf{X_{SG_g}}$. Moreover, $\Delta |\mathbf{V}_i|$ for node $i$ due to the DNR can be estimated as:

$$(|\mathbf{V}_{0i}| + \Delta |\mathbf{V}_i|)^2 = (V_{di0} + \Delta V_{di})^2 + (V_{qi0} + \Delta V_{qi})^2, \quad (15)$$

$$\text{where} \quad \Delta |\mathbf{V}_i| \approx |\mathbf{V}_{0i}|^{-1} \cdot (V_{di0} \cdot \Delta V_{di} + V_{qi0} \cdot \Delta V_{qi}),$$

$$= |\mathbf{V}_{0i}|^{-1} \cdot \mathbf{V}_{0i}^T \cdot \mathbf{V}_i = \mathbf{M}_i \cdot \mathbf{V}_i. \quad (16)$$

For all nodes $i = 1, \cdots, N$, the aggregated form of (16) is

$$\Delta |\mathbf{V}| = \mathbf{M} \cdot \Delta \mathbf{V}, \quad (17)$$

where $\mathbf{M} = \text{diag}(\mathbf{M}_1, \mathbf{M}_2, \cdots, \mathbf{M}_N)$. Using (11) and (17), $\Delta |\mathbf{V}|$ for $\Delta \mathbf{I_T}$ can be represented as:

$$\Delta |\mathbf{V}| = \mathbf{M} \cdot \mathbf{Z} \cdot (\mathbf{C_{SG}} \cdot \Delta \mathbf{X} + \Delta \mathbf{I_T}). \quad (18)$$

In compact form, (13) and (18) are given as:

$$\begin{bmatrix}\Delta f\\ \Delta |\mathbf{V}|\end{bmatrix} = \begin{bmatrix}\mathbf{C_F}\\ \mathbf{C_V}\end{bmatrix} \cdot \Delta \mathbf{X} + \begin{bmatrix}\mathbf{O}_{1 \times 2N}\\ \mathbf{D_V}\end{bmatrix} \cdot \Delta \mathbf{I_T}, \quad (19)$$

where $\mathbf{C_F} = \mathbf{H} \cdot \mathbf{S}$, $\mathbf{C_V} = \mathbf{M} \cdot \mathbf{Z} \cdot \mathbf{C_{SG}}$, and $\mathbf{D_V} = \mathbf{M} \cdot \mathbf{Z}$. The complete analytical model consists of (12) and (19), enabling estimation of $\Delta f$ and $\Delta |\mathbf{V}|$ for the DNR.

## III. CASE STUDIES AND SIMULATION RESULTS

The proposed analytical model was tested on the IEEE 37-node test feeder, as shown in Fig. 4, with modification of the line connections. The test bed operated as an islanded microgrid, where five SGs were used to regulate frequency and voltages. The total rated power of the SGs was set to 4.8 MVA and the detailed parameters are provided in [4]. The total load demand was set to $4.3 + j2.7$ MVA or, equivalently, $0.89 + j0.56$ pu. For simplicity, the load demand was assumed to be equally distributed among all nodes, and the ZIP coefficients

$$\Delta \mathbf{I_T} = \begin{bmatrix}\mathbf{O}_{1 \times 2(i-1)}, & \underbrace{\{\mathbf{y}_{ij}(\mathbf{V}_{0i} - \mathbf{V}_{0j})\}^T}_{= \Delta \mathbf{I_{T_i}}^T}, & \underbrace{\{\mathbf{y}_{ij}(\mathbf{V}_{0j} - \mathbf{V}_{0i}) - \mathbf{y}_{jk}(\mathbf{V}_{0j} - \mathbf{V}_{0k})\}^T}_{= \Delta \mathbf{I_{T_j}}^T}, & \underbrace{\{-\mathbf{y}_{jk}(\mathbf{V}_{0k} - \mathbf{V}_{0j})\}^T}_{= \Delta \mathbf{I_{T_k}}^T}, & \mathbf{O}_{1 \times 2(N-k)}\end{bmatrix}^T \quad (6)$$

for the load at each node were set to 0.5, 0.3, and 0.2, respectively [8]. Moreover, 3-*ph* balanced impedances were adopted for the distribution lines, based on the average values of each configuration [3], [9]. For the DNR, six switches were located arbitrarily throughout the network. Specifically, at $t = 1$ s, $SW_{1, 3, 4}$ were closed, and $SW_2$ was opened, to change the grid topology. At $t = 21$ s, $SW_{5, 6}$ were then closed to restore the de-energized load demand of $0.30 + j0.19$ pu in the areas $LA_1$ and $LA_2$. A comprehensive model of the test bed was implemented using MATLAB/SIMULINK, providing actual profiles of the frequency and voltage responses to the DNR.

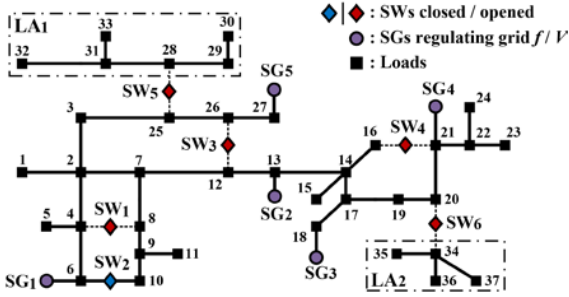

Fig. 4. Single-line diagram of the modified IEEE 37-node test feeder.

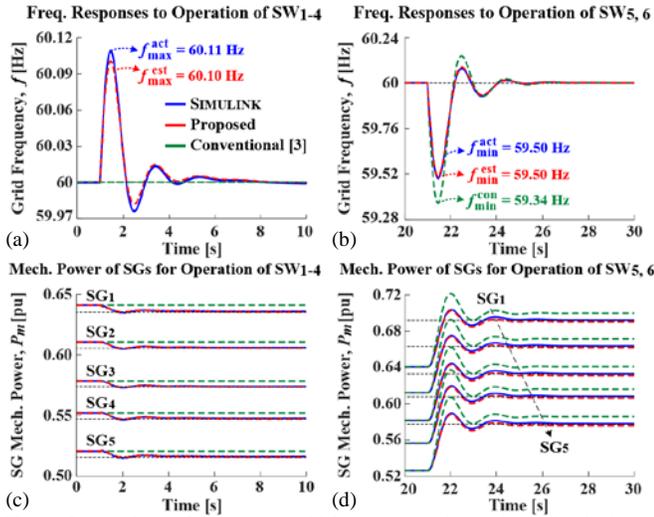

Fig. 5. Comparison of the (a), (b) grid frequency and (c), (d) mechanical power of the SGs with operation of $SW_{1-4}$ and $SW_{5, 6}$.

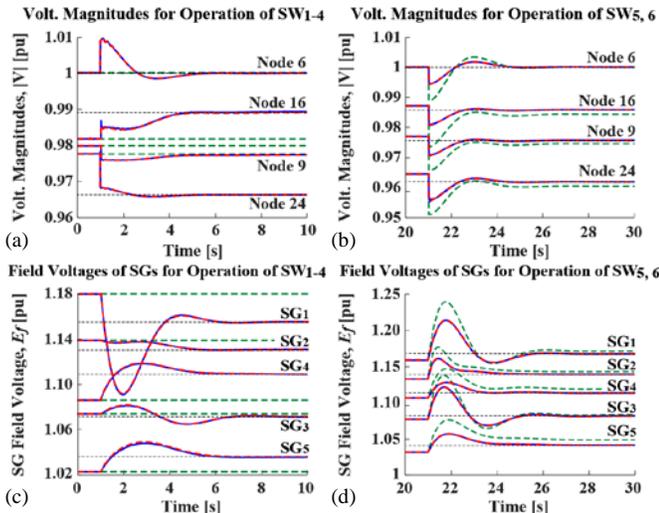

Fig. 6. Comparison of the (a), (b) node voltages and (c), (d) field voltages of the SGs with operation of $SW_{1-4}$ and $SW_{5, 6}$.

Fig. 5 compares the profiles of the grid frequency and the mechanical power of the SGs among the proposed and conventional analytical models and the comprehensive SIMULINK model. As shown in Fig. 5(a) and (c), the profiles for the proposed model were very similar to those of the SIMULINK model in both transient and steady states. This validates that the proposed model successfully reflects the responses of the SGs and loads, and the variations in the network power loss with operation of $SW_{1-4}$. Fig. 5(b) and (d) also show the good consistency between the proposed analytical model and the SIMULINK model, particularly when the grid experienced relatively large variations in frequency and mechanical power due to load restoration via the DNR.

Similarly, Fig. 6 compares the node and field voltages between the analytical and SIMULINK models. Fig. 6(a) and (b) show small differences in node voltage magnitude between the proposed model and the SIMULINK model at arbitrarily selected nodes. For all nodes, the average and maximum values of root mean square error (RMSE) (i.e., $\{(1/T)\cdot\Sigma_t(|\Delta\mathbf{V}_i^{est}|_t - |\Delta\mathbf{V}_i^{act}|_t)^2\}^{1/2}$) were estimated as $1.3 \times 10^{-4}$ pu and $9.5 \times 10^{-4}$ pu, respectively. The small RMSEs validate the proposed model. Moreover, Fig. 6(c) and (d) show good consistency in the field voltage responses of the SGs to the DNR between the proposed and SIMULINK models.

IV. CONCLUSION

This paper presents a new analytical model to estimate the dynamic responses of grid frequency and node voltages to the DNR. The switching operations were modelled as discrete variations in line admittances and, consequently, step variations in the node-injected currents. The proposed analytical model was then established by integrating the network model, characterized by the step variations, with the dynamic models of the SGs and voltage-dependent loads. The results of the comparative case studies validate that the proposed analytical model successfully reflected the dynamic variations in frequency, node voltages, and SG responses for the DNR, with or without load restoration.